\pdfoutput=1
\RequirePackage{ifpdf}
\ifpdf
  \documentclass[10pt,a4paper,pdftex]{article}
\else
  \documentclass[10pt,a4paper,dvips]{article}
\fi

\ifpdf
  \usepackage{color}
\fi

\usepackage[latin1]{inputenc}
\usepackage{amsmath}
\usepackage{amsfonts}
\usepackage{amssymb}
\usepackage{float}
\usepackage{epsfig}
\usepackage{subfigure}
\usepackage[section]{placeins}
\usepackage{pstricks}
\usepackage{mathrsfs}

\ifpdf
  \usepackage{graphicx}
  \usepackage[unicode]{hyperref}
\else
  \usepackage{graphics}
\fi

\newcommand{\td}{\mathrm{d}}

\allowdisplaybreaks[1]

\author{Johannes Schmude\footnote{pyjs@swansea.ac.uk}\\
  \mbox{}\\
  Department of Physics\\
  Swansea University, Swansea, SA2 8PP, United Kingdom}

\title{Comments on the distinction between color- and flavor-branes and
  new D3-D7 solutions with eight supercharges}

\date{}

\begin{document}

\maketitle

\begin{abstract}
We investigate the distinction between color- and flavor-branes that
is usually made in the context of gauge/string duality with
backreacting flavors. Our remarks are based on a series of examples
concerning the role of source terms in relatively simple supergravity
backgrounds that allow for a well-controlled approach to the
problem. The observations suggest that, in opposite to general
practice, one could consider such terms for both kinds of branes,
while their presence is only essential for smeared sources -- as is
commonly the case for flavor-branes.

Among the examples studied are D$3$-D$7$ systems with eight
supercharges, where the D$7$-branes are assumed to be
smeared. Starting from a fairly generic ansatz, we will find new
analytic and numeric solutions and briefly compare these to previous
work in this field.
\end{abstract}

\section{Introduction}
\label{sec:introduction}

In the context of gauge/string duality, recent years have seen the
adoption of a standard method when it comes to the study of gauge
theories with a large number of flavors in the Veneziano
limit using supergravity- and brane-actions. Starting with
\cite{Bigazzi:2005md}, \cite{Casero:2006pt} and reviewed in
\cite{Nunez:2010sf}, the methodology is founded on the addition of
further branes to the background, that are space-time filling while
also extending along non-compact transverse cycles.

Prior to flavoring, the supergravity background is always thought of
as the near-horizon geometry of a stack of branes that might wrap a
compact cycle \cite{Maldacena:2000yy} or be placed at the tip of a
singular manifold \cite{Klebanov:1999tb}, but is in each case
described by the equations of motion of the suitable ten- or
eleven-dimensional supergravity. That is, there is an action
$S_{\text{IIA/B}}$ or $S_{\text{M}}$ that gives rise to the relevant equations of
motion solved by the background in question, which is argued to be
dual to a certain gauge theory.

Now, to add flavors to the system, one adds to this supergravity
action the action of the new flavor-branes $S_{\text{flavor}}$, and
solves the equations of motion of the system
\begin{equation}\label{eq:flavor_action}
  \begin{aligned}
    S &= S_{\text{M,IIA/B}} + S_{\text{flavor}}
  \end{aligned}
\end{equation}
$S_{\text{flavor}}$ is a superposition of $N_f$ D-brane actions
consisting of the standard DBI- and Wess-Zumino terms. There is a
multitude of arguments supporting this procedure, some of which we
will briefly mention here. See \cite{Nunez:2010sf} for a more
thorough discussion. From the point of view of the 't Hooft expansion
of the gauge theory, flavor should lead to diagrams with boundaries,
corresponding to an open string sector in the dual string theory. And
as it is well known, one adds an open string sector to a type II
theory by including D-branes. Apart from the fact that only the flavor
branes appear in the action (\ref{eq:flavor_action}) there are further
conceptual differences between the color and flavor-branes in the
background. Flavor is a global symmetry while color is related to a
local, gauged one. Hence flavor charged objects exist in non singlet
states, while this is only possible for color-charged ones in
non-confining phases.

The difference between the two kinds of branes mentioned is reflected
in (\ref{eq:flavor_action}) -- there is a source term for the flavor
branes, while the physics of the color-branes is captured by the
supergravity action. From the point of view of gauge/string duality,
the physics of the pure Yang-Mills sector (e.g.~glueballs) are
captured by the supergravity action, those of the open strings
describing the fundamental matter (mesons) by the brane action and
interactions between the two by the fact that the background fields as
well as world-volume fields couple in the brane action. In this paper,
we will critically investigate this statement, the form of
(\ref{eq:flavor_action}) and the distinction between flavor- and
color-branes. Working in the supergravity limit, our observations will
be based on a series of examples signifying the relevance of
source-terms such as $S_{\text{flavor}}$ in (\ref{eq:flavor_action})
for various brane-solutions.

In section \ref{sec:p-brane-sources} we will recall some known yet often
overlooked\footnote{
See however \cite{Bertolini:2000dk} and \cite{Bertolini:2001qa} which,
working in the boundary state formalism, do include source terms for
localized color- and flavor-branes.}
facts about source terms for $p$-branes in supergravity -- 
we are thinking here of the standard $1/2$-BPS brane solutions of type
IIA/B and M-theory. These are usually found by studying the equations
of motion of a supergravity action $S_{\text{M,IIA/B}}$. Yet as we
will remind the reader, the $p$-brane solutions solve these equations
of motion only if one excises the locations of the branes from
space-time. Upon adding a suitable source term to the equations of
motion the equations of motion are solved everywhere. From the point
of view of finding solutions to the equations of motion, one can
ignore the source term as the sources are not distributed over an open
subset of space-time; they can be smeared along some directions, as
long as they are localized in others.

This changes in section \ref{sec:smeared-sources} where we will see
that source terms are essential once we start to smear the branes over
open subsets. The crucial point is that sources generally lead to the
violation of the Bianchi identity of a magnetic field strength,
e.g.~$\td F_{(D-p-2)} = \rho_{(D-p-1)}$. Only as long as the branes
are localized can one ignore the source-density $\rho$ and proceed by
working with the supergravity action alone.\footnote{We do simplify
  things here, as Bianchi identities can also be violated by the
  presence of Chern-Simons terms. In most of the examples we have in
  mind however, this is not the case.}

These observations of sections \ref{sec:p-brane-sources} and
\ref{sec:smeared-sources} imply that in the generic case
(\ref{eq:flavor_action}) implicitly contains a source-term for the
color-branes and should be replaced with
\begin{equation}
  \label{eq:flavor_color_action}
  S = S_{\text{M,IIA/B}} + S_{\text{color}} + S_{\text{flavor}}
\end{equation}
With this in mind, we will turn in section \ref{sec:d3d7-solutions} to
the relatively simple problem of D$3$-D$7$ systems with eight
supercharges, corresponding to duals of $\mathcal{N}=2$ gauge theories
in $d=3+1$ dimensions. Here we will compare the backgrounds found by
\cite{Aharony:1998xz} - \nocite{Grana:2001xn} \cite{Kirsch:2005uy},
which solve the pure supergravity equations of
motion without any source terms for either flavor or color-branes, to
new solutions, found working in the spirit of \cite{Casero:2006pt}
solving the system given by $S_{\text{IIB}} + S_{\text{flavor}}$. As
we will see, the two approaches lead to a set of equations and
solutions which only differ by the fact that our new ansatz implies
that the D$7$-branes are smeared. From this it can be implied that the
two approaches are equivalent. Moreover, we will see that a series of
T-dualities can exchange the color- and flavor-branes, although we did
include source terms for the latter and not for the former. From all
these observations we conclude that equation
(\ref{eq:flavor_color_action}) is the most suitable ansatz, leading us
to the one of the main observation of this paper: From a technical
point of view, source terms can be ignored as long as the sources are
localized, yet when using (\ref{eq:flavor_action}) to make statements
about the dynamics of the resulting system, care has to be taken.

Finally, we use the results of section \ref{sec:d3d7-solutions} to
find some new analytic (numeric) solutions to the D$3$-D$7$ system in
section \ref{sec:analytic-solutions}
(\ref{sec:numeric-solutions}). Note that many of the results presented
in section \ref{sec:p-brane-sources} are not new, but seem to be often
overlooked in this line of research.

\section{$p$-brane sources}
\label{sec:p-brane-sources}

As a warm-up, we will review source terms for $p$-brane solutions (see
\cite{Ortin:2004ms}). The equations of motion derived from\footnote{
As we mentioned in the introduction, our ansatz here excludes the
possibility of Chern-Simons terms as are relevant for
\cite{Klebanov:2000hb} for example.}
\begin{equation}
  \label{eq:p-brane_a-model_sugra-action}
  \begin{aligned}
    S_{\text{grav}} &= \frac{1}{16 \pi G_{D}} \int \lbrack \td^D x
      \sqrt{-g} (R - \frac{1}{2} \partial^\mu \Phi \partial_\mu \Phi)
      - \frac{1}{2} e^{a \Phi} F_{(p+2)} \wedge *F_{(p+2)} \rbrack
  \end{aligned}
\end{equation}
are solved by
\begin{equation}
  \label{eq:p-brane_ansatz}
  \begin{aligned}
    \td s^2 &= H_d(y)^{-2\frac{d-2}{\Delta}} \td x_{1,p}^2 +
    H_d(y)^{2\frac{p+1}{\Delta}} \td y_d^2 \\
    F_{(p+2)} &= e^{-\frac{a}{2}\Phi_\infty} \sqrt{2 \frac{D-2}{\Delta}}\td
    (H_d(y)^{-1} - 1) \wedge 
    \td x^0 \wedge \dots \wedge \td x^{p} \\
    e^\Phi &= e^{\Phi_\infty} H_d(y)^{a \frac{D-2}{\Delta}}
  \end{aligned}
\end{equation}
with $\Phi_\infty$ constant, $D = (p + 1) + d$, $y^2 = y^a y^a$ and
\begin{equation}
  \label{eq:p-brane_ansatz_definitions}
  \begin{aligned}
    \Delta &= (p+1)(d-2) + \frac{1}{2} a^2 (D-2) \\
    H_d &= 1 + \left\{
    \begin{array}{cc}
      h_1 | y | & d = 1 \\
      h_2 \log y & d = 2 \\
      \frac{h_d}{y^{d-2}} & d \geq 3
    \end{array}\right.
  \end{aligned}
\end{equation}
the standard $1/2$-BPS $p$-brane solutions of supergravity.
The solutions are interpeted as describing branes at
$y=0$. However, (\ref{eq:p-brane_ansatz}) do not solve the equations
of motion \emph{everywhere}, as $H_d(y)$ satisfies
\begin{equation}
  \label{eq:fundamental_solution_of_Laplace-equation}
  \begin{aligned}
    \square_{\mathbb{R}^d} H_d &= - \frac{d(d-2) \pi^{d/2}
      h_d}{\Gamma(d/2)} \delta^{(d)}(y)
  \end{aligned}
\end{equation}
and the equations of motion do contain the $d$-dimensional Laplacian
$\square_{\mathbb{R}^d}$. These singularities are of course due to the
$p$-brane at the origin and can be lifted by adding a source term to
the action, $S = S_{\text{grav}} + S_{\text{src}}$,
\begin{equation}
  \label{eq:brane_action_with_auxiliary_metric}
  \begin{aligned}
    S_{\text{src}} &= -\frac{T_p}{2} \int \td^{p+1}\xi \sqrt{-\gamma}
    e^{b\Phi} \lbrack \gamma^{ij} \partial_i
    X^\mu \partial_j X^\nu g_{\mu\nu} + (p - 1) \rbrack + \mu_p \int
    X^*C_{(p+1)}
  \end{aligned}
\end{equation}

$S_{\text{src}}$ introduces additional terms to the equations of
motion.\footnote{
Note that some of the mathematical manipulations in the following
section will be quite cavalier. Matching singularities, we will
perform calculations with $\delta$-functions without carefully
regulating these.}
E.g.~in the case of the Maxwell equation
\begin{equation}
  \label{eq:source_modified_maxwell_equation}
  \begin{aligned}
    0 &= \partial_\mu (\sqrt{-g} e^{a\Phi} F^{\mu \nu_0 \dots
      \nu_{p}}) \\
    &+16 \pi G_{D} \mu_p \int \td^{p+1}\xi \epsilon^{i_0 \dots
      i_p} \partial_{i_0} X^{\nu_0} \dots \partial_{i_p} X^{\nu_p}
    \delta^{(D)}(x-X(\xi))
  \end{aligned}
\end{equation}
which can be rewritten as $\td (* e^{a\Phi} F) \sim
\delta^{(d)}(y)$. So the presence of the source leads to the violation
of the Bianchi identity of the magnetically dual field strength. In
static gauge\footnote{In static gauge
  \begin{equation*}
      X^\mu(\xi^i) = \left\{
      \begin{array}{cc}
        \xi^\mu & \mu \leq p \\
        0 & \mu > p
      \end{array} \right.
  \end{equation*}
}
these can be easily seen to be localized at $y = 0$. They match the
singularities arising from $\square_{\mathbb{R}^d} H$ if
\begin{equation}
  \label{eq:constants_fixed_by_source_term}
  \begin{aligned}
    h_d &= \frac{16\pi G_{D} T_p e^{-\frac{a}{2}\Phi_\infty}\Gamma(d/2)}{d
      (d-2)\pi^{d/2}} \frac{\Delta}{2(D-2)} \\
    \frac{T_p}{\mu_p} &= \sqrt{2\frac{D-2}{\Delta}} e^{a\Phi_\infty}
    \\
    b &= -\frac{a}{2}
  \end{aligned}
\end{equation}

It is worthwhile to take a closer look at the equation of motion for
the embedding fields $X^\mu(\xi)$. As no fields in our ansatz depend
on the world-volume coordinates $\xi^i$, the Euler-Lagrange
equations $\partial_i \frac{\partial \mathcal{L}}{\partial \partial_i
  X^\mu} - \frac{\partial \mathcal{L}}{\partial X^\mu}$ reduce to
\begin{equation}
  \label{eq:world-volume_X_equation}
  \begin{aligned}
    \frac{\partial \mathcal{L}}{\partial X^\mu} &= \frac{T_p}{2}
    \sqrt{-\gamma} e^{b\Phi} \lbrack b \partial_\mu \Phi
    (\gamma^{ij} \partial_i X^\kappa \partial_j X^\lambda - p +1 )
    + \gamma^{ij} \partial_i X^\kappa \partial_j
    X^\lambda \partial_\mu g_{\kappa\lambda} \rbrack \\
    &- \frac{\mu_p}{(p+1)!} \epsilon^{i_0 \dots i_p} \partial_{i_0}
    X^{\mu_0} \dots \partial_{i_p} X^{\mu_p} \partial_\mu C_{\mu_0
      \dots \mu_p}
  \end{aligned}
\end{equation}
which vanishes identically for $\mu \in \{0, \dots, p\}$. Let us
however generalize this part of the discussion to include non-extremal
$p$-branes. That is, we assume the metric to take the form
\begin{equation}\label{eq:non-extremal_metric}
  \td s^2 = H_d(y)^{-2\frac{d-2}{\Delta}} \lbrack - f(y) \td t^2 + \td
  x_p^2 \rbrack
  + \dots
\end{equation}
where we dropped the transverse directions, which include off-diagonal
elements in our choice of coordinates, but do not contribute to the
following discussion. (\ref{eq:world-volume_X_equation}) reduces to
\begin{equation}
  \begin{aligned}
    0 &= \lbrack \frac{1}{2} a^2 \frac{D-2}{\Delta} +
    \frac{(p+1)(d-2)}{\Delta} f \rbrack H_d^{-2} \partial_\mu H_d -
    \frac{p+1}{2} H_d^{-1} \partial_\mu f
  \end{aligned}
\end{equation}
Setting $f = 1 - \frac{Q}{y^{d-2}}$, it follows that the above is only
solved if $Q = 0$, i.e.~if the brane is extremal. Note further that in
the non-extremal case, the term $H_d^{-1} \partial_\mu f$ diverges as $y
\to 0$, while $H_d^{-2} \partial_\mu H_d \to 0$. One can interpret this
behavior in the light of supersymmetry. By introducing $f$, we only
modify the part of the brane action coupling to the metric, but not
the one coupling to the $p+2$-form. The $X^\mu$ equation of motion can
be thought of as a balancing between these two sectors (it imposes a
relation between $T_p$ and $\mu_p$), so it is no surprise that it
holds no longer once we have perturbed this balance. This might simply
indicate an instability of the embedding or might indicate that it is
not possible to find a source term for the non-extremal
solution.\footnote{The author is not aware of any   general theorems
  regarding the existence of source terms for classical theories of
  gravity.}
One should take in the account \cite{Bigazzi:2009bk}, where
the authors constructed a finite temperature background including
flavor-branes. In opposite to our discussion in the previous paragraph
however, this background's non-extrmality is due to a horizon
associated with the color-branes, while only the flavor-branes are
represented by a source.

Dropping the $1$ in the harmonic functions $H_d(y)$ in
(\ref{eq:p-brane_ansatz_definitions}) leads generally to the
near-horizon limit of the extremal $p$-brane considered. From
(\ref{eq:fundamental_solution_of_Laplace-equation}) it follows however
that the source-terms are still necessary in this limit -- the
argument does not depend on the asymptotic value of
$H_d$.\footnote{After all, no matter whether in the near-horizon limit
  or not, $H_d$ is harmonic everywhere except at the origin. See
  e.g.~chapter 2.2 of \cite{Evans:1998}.}
Hence one can argue that to
fully solve the equations of motion, one should add the source term
for both the $p$-brane solutions as well as their near-horizon limit.

One should note that in this context, the introduction of the source
term (\ref{eq:brane_action_with_auxiliary_metric}) did not add
additional degrees of freedom to the background -- $\gamma_{ij}$ is
auxiliary and $X^\mu(\xi)$ is merely a choice of gauge -- but allowed
us to lift the singularities as well as impose some relations between
$h_d, T_p, \mu_p, G_{D}$. In other words, by matching the source term
with the $p$-brane solutions, we can give fix the charge and tension
$\mu_p, T_p$ of the brane. Neither was the source term necessary to
find the solution -- we could have worked immediately with
$S_{\text{grav}} + S_{\text{src}}$, but there was no need to do so.

\section{Smeared sources}
\label{sec:smeared-sources}

The situation changes a bit when we consider smeared sources. That is,
the brane sources are not taken to be localized, but are continuously
distributed over some open subset of space-time. Smearing was first
seen in the context of T-duality and has today seen widepread use in
the the field of gauge/string duality with a large number of flavors,
$N_f \sim N_c$ (\cite{Casero:2006pt}, \cite{Nunez:2010sf},
\cite{Casero:2007jj}- \nocite{HoyosBadajoz:2008fw}
\nocite{Benini:2006hh} \nocite{Bigazzi:2008qq}
\nocite{Gaillard:2008wt} \nocite{Gaillard:2009kz}
\nocite{Benini:2007gx} \nocite{Bigazzi:2008zt}
\cite{Arean:2008az}). Here it simplifies the search for solutions as
while also avoiding the problem of including corrections to the DBI-
and Wess-Zumino terms appearing in the flavor action.

To allow for smearing, we include a distribution density $\rho_{(d)}$ in the
source term, which is formally a $d$-form on the space transverse to
the additional branes. Introducing also a calibration form
$\mathcal{K}_{(p+1)}$, which is essentially a volume form for the
brane, one can then write the source term for \emph{supersymmetric}
sources as (see \cite{Gaillard:2008wt} for details)
\begin{equation}
  \label{eq:calibrated_brane_action}
  \begin{aligned}
    S_{\text{src}} &= -T_p \int (e^{b\Phi} \mathcal{K}_{(p+1)} -
    C_{(p+1)}) \wedge \rho_{(d)}
  \end{aligned}
\end{equation}
Calculating the resulting equations of motion, the Maxwell equation
takes the form
\begin{equation}
  \label{eq:source_modified_maxwell-equation}
  \td (*e^{a\Phi} F_{(p+2)}) = 16\pi G_{D} T_p \rho_{(d)}
\end{equation}
-- a straightforward generalization of
(\ref{eq:source_modified_maxwell_equation}). As a matter of fact, as
long as there is some supersymmetry, it is sufficient to study the
form-field equations such as
(\ref{eq:source_modified_maxwell-equation}) together with the
supersymmetry conditions. The Einstein and Dilaton equations are then
implied \cite{Koerber:2007hd}. In contrast to the localized case of
section \ref{sec:p-brane-sources}, we would not have been able to
derive suitable equations of motion without the source term, so in the
context of smearing (over an open subset), the source term is
essential.

\section{D$3$D$7$ solutions}
\label{sec:d3d7-solutions}

With all this in mind, let us take a look at D$3$-D$7$ solutions with
$8$ supercharges. This has previously been studied in
\cite{Bertolini:2000dk} -\nocite{Bertolini:2001qa}
\nocite{Aharony:1998xz} \nocite{Grana:2001xn} 
\cite{Kirsch:2005uy}
in the case where the D$7$-branes are localized. Note that the authors
of \cite{Aharony:1998xz} - \nocite{Grana:2001xn} \cite{Kirsch:2005uy}
did not include any source terms in their actions working with $S =
S_{\text{IIB}}$, while \cite{Bertolini:2000dk} and
\cite{Bertolini:2001qa} do include source terms for color- and
flavor-branes. From our remarks in section \ref{sec:p-brane-sources}
we suspect that this is not necessary (as their sources are
localized), but we will see so explicitly. First, let us briefly
summarize the background of \cite{Kirsch:2005uy} (in string frame):
\begin{equation}
  \label{eq:polchinskin-grana-vaman-kirsch}
  \begin{aligned}
    \td s^2 &= H^{-1/2} \td x^2 + H^{1/2} ( \td z_1 \td \bar{z}_1 +
    \td z_2 \td \bar{z}_2 + e^{\Psi(z_3, \bar{z}_3)} \td z_3 \td
    \bar{z}_3 ) \\
    e^{\Psi(z_3, \bar{z}_3)} &= \tau_2(z_3) |\eta(\tau)|^4
    |z_3|^{-N_f/6} \\
    \tau &= C_{(0)} + \imath e^{-\Phi} \\
    F_{(5)} &= -\frac{1}{2\sqrt{2} (2\pi)^{7/2} g_s (\alpha^\prime)^2}
    (1+*) \td H^{-1} \wedge \td x^0 \wedge \dots \wedge \td x^3
  \end{aligned}
\end{equation}
The complex structure (or axio-dilaton) $\tau$ is fixed by the
presence of \emph{localized} D$7$ branes. Crucial for us is that the
warp factor $H(z_i, \bar{z}_i)$ must satisfy a deformation of the
Laplace equation on the transverse space,
\begin{equation}
  \label{eq:PGVK_Laplace-equation}
  ( \partial_1 \bar{\partial}_1 + \partial_2 \bar{\partial}_2 +
  e^{-\Psi} \partial_3 \bar{\partial}_3 ) H = 0
\end{equation}
Strictly speaking, we are not interested in solutions to a modified Laplace
equation, but a modified Poisson equation, as D$3$-branes will appear
in a singularity, just as in the $p$-brane case
\begin{equation}
  \label{eq:PGVK_Poisson-equation}
  ( \partial_1 \bar{\partial}_1 + \partial_2 \bar{\partial}_2 +
  e^{-\Psi} \partial_3 \bar{\partial}_3 ) H = \delta^{(6)}(z)
\end{equation}
In the following we will encounter different examples of $H$ with
different kinds of $\delta$-functions appearing on the right hand side
of equations like (\ref{eq:PGVK_Poisson-equation}). To simplify the
notation, we shall always drop the $\delta$-function, write the
equations as (\ref{eq:PGVK_Laplace-equation}) but keep in mind that
$H$ is usually singular at some point.

Looking for new solutions and working in the spirit of the flavoring
program, we study $S = S_{\text{IIB}} + S_{\text{flavor}}$ with the
source term being a superposition of D$7$ actions. Then we make the
Ansatz (Einstein frame)
\begin{equation}
  \label{eq:ansatz}
  \begin{aligned}
    \td s^2 &= e^{-\frac{1}{2}\Phi} \lbrack e^{2f} \td x_{1,3}^2 +
    e^{2g} \td v_4^2 + e^{2h} ( \td w^2 + w^2 \td \phi^2 ) \rbrack \\
    F_{(5)} &= (1+*_{10}) ( \td f_5 \wedge \td x^{0123} ) \\
    F_{(1)} &= f_1(w) w \td \phi \\
    \Phi &= \Phi (v,w)
  \end{aligned}
\end{equation}
Where $f, g, h, f_5, \Phi$ depend on $w, v = \sqrt{v^i v^i}$ while
$f_1$ depends on $w$ alone. The most striking difference between
(\ref{eq:ansatz}) and (\ref{eq:polchinskin-grana-vaman-kirsch}) is
that our choice for $F_{(1)}$ is in general not exact and can thus not
be understood in terms of a $0$-form potential $C_{(0)}$ and the
relation $F_{(1)} = \td C_{(0)}$. In contrast, the appearance of
$C_{(0)}$ in (\ref{eq:polchinskin-grana-vaman-kirsch}) implies $\td
F_{(1)} = 0$, except at isolated singularities.\footnote{
The non-exactness of $F_{(1)}$ explains also why in opposite to the
earlier papers we do not rely on holomorphy of the axio-dilaton in the
$(w,\phi)$ plane. If $F_{(1)}$ is exact, the supergravity variations
can be phrased in terms of $C_{(0)}$ and the dilatino variation
quickly takes the form of Cauchy-Riemann equations for $e^{-\Phi} +
\imath C_{(0)}$.}
This is why the former ansatz will not allow for smeared D$7$
branes. Of course we study the action $S_{\text{IIB}} +
S_{\text{src}}$, so there will be $\rho_{(2)}$ such that $\td F_{(1)}
= \rho_{(2)}$. In other words, we will not need to impose a Bianchi
identity for $F_{(1)}$, but are on the contrary rather interested in
its explicit violation. Note also that our choice for $F_{(1)}$
implies that all D$7$ sources will be smeared along $\phi$.

Demanding the existence of a SUSY spinor $\epsilon$ satisfying $\imath
\Gamma^{0123} \epsilon = -\epsilon$ and $\Gamma^{4567} \epsilon =
\epsilon$ we study the BPS-system given by
\begin{equation}
  \label{eq:SUGRA_equations}
  \begin{aligned}
    0 &\overset{!}{=} \delta_\epsilon \lambda = \frac{1}{2}
    (\partial_\mu \Phi - \imath e^{\Phi} F_\mu) \Gamma^\mu \epsilon \\
    0 &\overset{!}{=} \delta_\epsilon \psi_\mu = \partial_\mu \epsilon
    + \frac{1}{4} \omega_{\mu ab} \Gamma^{ab} \epsilon + \frac{\imath}{4}
    e^\Phi F_\mu + \frac{\imath}{16} \frac{1}{5!}
    F_{\nu\rho\sigma\tau\upsilon} \Gamma^{\nu\rho\sigma\tau\upsilon}
    \Gamma_\mu \epsilon
  \end{aligned}
\end{equation}
as well as the Bianchi identity for $F_{(5)}$. As we mentioned
earlier, integrability ensures that the remaining equations of motion
will be satisfied. One then sees quickly that any solution of the
original ansatz can be rewritten in terms of only two functions,
$H(v,w), \Delta_{gf}(w)$, and a set of integration constants
\begin{equation}\label{eq:analytic_metric}
  \begin{aligned}
    \td s^2 &= e^{-\frac{c_\Phi}{2}} \{ H^{-1/2} \td
    x_{1,3}^2 + H^{1/2} \lbrack \td v_4^2 + e^{-2
      (\Delta_{gf} - c_h)} (\td w^2 + w^2 \td \phi^2) \rbrack \} \\
    F_{(5)} &= (1+*) \td \lbrack ( e^{- 2 c_\Phi} H^{-1} +
    c_{f_5} ) \td x^0 \wedge \td x^1 \wedge \td x^2 \wedge \td x^3
    \rbrack \\
    F_{(1)} &= w (\partial_w e^{-2\Delta_{gf}}) \td \phi \\
    \Phi &= 2\Delta_{gf} + c_\Phi
  \end{aligned}
\end{equation}
subject to the modified Laplace/Poisson
equation\footnote{\label{fn:comments_about_F5_bianchi} Crucially,
  (\ref{eq:elliptic_PDE}) arises from the Bianchi identity on
  $\td F_{(5)} = 0$. As we have seen in sections
  \ref{sec:p-brane-sources} and \ref{sec:smeared-sources}, these
  identities relate directly to the presence of sources and should be
  rewritten as $\td F_{(5)} = \rho_{(6)}$ as we are looking for
  backgrounds with D$3$ sources. So strictly speaking, there should be
  a source density on the right hand side of
  (\ref{eq:elliptic_PDE}), at least a $\delta$-function. As we are
  looking for smeared D$7$ branes in backgrounds with localized D$3$s, we ignore
  this distinction and just keep in mind that when solving
  (\ref{eq:elliptic_PDE}), we are looking for solutions that show
  singular behavior at $(v,w) = (0,0)$.}
\begin{equation}
  \label{eq:elliptic_PDE}
  \begin{aligned}
    0 &= \left\lbrack ( \partial_v^2 +
        \frac{3}{v} \partial_v ) + e^{\Delta_{gf}(w)-c_h} (
        \partial_w^2 + \frac{1}{w} \partial_w ) \right\rbrack H(v,w)
  \end{aligned}
\end{equation}
which can be more succinctly summarized as
\begin{equation}
  0 = ( \square_v + e^{\Delta_{gf}(w) - c_h} \square_w ) H(v,w)
\end{equation}
Apart from the $w$ and $z, \bar{z}$ dependence, this is the same
equation as (\ref{eq:PGVK_Laplace-equation}). However, while
(\ref{eq:polchinskin-grana-vaman-kirsch}) was derived without use of
an additional source term, the derivation of (\ref{eq:elliptic_PDE})
was based on $S_{\text{IIB}} + S_{\text{D}7}$. As we found previously,
as long as the sources are localized, one is free not to include the
source term. Note that working in the spirit of gauge/string duality
with flavor, we did not include a source term for the D$3$
\emph{color} branes -- yet of course, we could have. See footnote
\ref{fn:comments_about_F5_bianchi}.

\subsection{An aside: T-dualities}
\label{sec:an-aside-t-duality}
It is instructive to take a look at various T-dualities. There are two
cases of interest -- performing four T-dualities along the $v^i$, or
performing two in the $(w,\phi)$ plane. In the latter case it is
appropriate to change coordinates to Cartesian ones -- $(w^1, w^2)$ --
to perform the dualities. The first case gives
\begin{equation}
  \label{eq:4T-dualities_along_v}
  \begin{aligned}
    \td s^2 &= e^{-\frac{c_\Phi}{2}} \{ e^{\Delta_{gf}} \td x_{1,3}^2
    + e^{-\Delta_{gf}} \lbrack \td v_4^2+ e^{2 c_h} H
    (\td w^2 + w^2 \td \phi^2) \rbrack \} \\
    \Phi &= c_\Phi - \log H \\
    F_{(5)} &= (1+*) \td ( e^{2\Delta_{gf}} \td x^0 \wedge \td x^1
    \wedge \td x^2 \wedge \td x^3 ) \\
    F_{(1)} &= - e^{-2c_\Phi} \partial_w (e^{-2c_\Phi} H + c_{f_5}) w
    \td\phi
  \end{aligned}
\end{equation}
Comparing \eqref{eq:analytic_metric} and
\eqref{eq:4T-dualities_along_v} shows the result of the dualitites to
be a swap $-2\Delta_{gf} \leftrightarrow \log H$. Now note that while the
Buscher rules for T-dualities in supergravity only apply for $v^i$ to
be an isometry of the background, i.e.~for $\partial_{v^i} H = 0$, the
substitution $-2\Delta_{gf} \leftrightarrow \log H$ is valid at the
level of the BPS equations and equations of motion too. The simple
reason is that the BPS equations are all trivially satisfied when
written in terms of $\Delta_{gf}$ and $H$, the equation of motion for
$F_{(1)}$ is always satisfied as well as $F_{(1)}$ depends only on
$\td \phi$, so the only points of interest are the Bianchi identities
for $F_{(5)}$ and $F_{(1)}$. These however do not need to be satisfied
if we allow for smeared brane sources. Again we point out that we only
included an explicit source term for the D$7$-branes, that have now
been turned into D$3$s.

T-dualities along $w^1, w^2$ lead to
\begin{equation}
  \label{eq:two_T-dualities_along_w}
  \begin{aligned}
    \td s^2 &= e^{-\frac{\Delta_{gf}}{2}-\frac{c_\Phi}{2}+c_h} \lbrack H^{-1/4} ( \td
    x_{1,3}^2 + e^{-2c_h} \td w_2^2) + H^{3/4} \td
    v_4^2 \rbrack \\
    \Phi &= 3\Delta_{gf} + c_\Phi - 2 c_h - \frac{1}{2}
    \log H \\
    F_{(7)} &= \td\lbrack (e^{-2c_\Phi}H^{-1} + c_{f_5}) w \td
    x^0 \wedge \dots \wedge \td x^3 \wedge \td w \wedge \td \phi
    \rbrack \\
    F_{(1)} &= -\td (e^{-2\Delta_{gf}(w)})
  \end{aligned}
\end{equation}

For $\Delta_{gf} = 0$, \eqref{eq:4T-dualities_along_v} and
\eqref{eq:two_T-dualities_along_w} reduce to the standard flat D$7$
and D$5$ solutions. This is of course expected, as \eqref{eq:ansatz}
describes a stack of D$3$-branes in flat space. Turning on
$\Delta_{gf}$ adds five- and one-form flux to the T-dual backgrounds
respectively; for (\ref{eq:two_T-dualities_along_w}) the one-form flux
is exact, however, so there are only additional D$7$ sources if
$\Delta_{gf}(w)$ is not differentiable at isolated points. We are
dealing with a D$7$-D$3$ and a D$5$(-D$7$) system, respectively.

In the context of gauge/string duality, the T-dualities along the
$v^i$ should be of interest. After all, it exchanges the $N_c$ color
D$3$-branes with the $N_f$ flavor D$7$-branes -- at first glance, we
have a duality $(N_c, N_f) \leftrightarrow (N_f, N_c)$. Of course the
precise form of the duality depends on the brane distributions.

\subsection{Simple, known solutions}
\label{sec:known-solutions}
For $\Delta_{gf} = 0, c_h = 0$, there is of course the standard D$3$-brane
solution,
\begin{equation}
  \label{eq:D3-brane_solution}
  \begin{aligned}
    H_3 &= 1 + \frac{r_3^4}{(v^2+w^2)^2}
  \end{aligned}
\end{equation}
the laplacian of which has a $\delta$-function singularity at
$(v,w)=(0,0)$ due to the presence of the D$3$-branes. The near horizon
limit is given by
\begin{equation}
  H_3 \mapsto \frac{r_3^4}{(v^2+w^2)^2}
\end{equation}

There are further solution that depend on only one variable and have
thus additional isometries in the background
\begin{equation}\label{eq:lesser_known_solutions}
  \begin{aligned}
    H(v,w) &= 1 + \frac{r_5^2}{v^2} \\
    H(v,w) &= 1 + r_7 \log w
  \end{aligned}
\end{equation}
They are the harmonic functions in four and two dimensions
respectively.\footnote{If one wonders why (\ref{eq:elliptic_PDE}) is
  not symmetric under $v \leftrightarrow w$ for $\Delta_{gf} = c_h$,
  the explanation can be found here. $v$ and $w$ are the radial
  coordinate in spaces of different dimension.}
They are singular at $v = 0$ or $w = 0$. The standard interpretation
here is to think of the D$3$-branes as having been smeared over
$(w,\phi)$ or the $v^i$. I.e.~the smeared branes are now codimensions
four or codimension two objects. Performing two (four) T-dualities
along the additional isometries leads to the standard D$5$ (D$7$) solutions.
Remeber that (\ref{eq:elliptic_PDE}) is linear, so any superposition
of (\ref{eq:D3-brane_solution}) and (\ref{eq:lesser_known_solutions})
is a solution as well.

\subsection{Analytic solutions}
\label{sec:analytic-solutions}

Looking for new solutions of (\ref{eq:elliptic_PDE}), we will make use
of the fact that there are not cross derivative terms of the form
$\partial_v \partial_w$. Hence the PDE is separable and we may look
for solutions of the form
\begin{equation}
  \begin{aligned}
    H(v,w) &= H_v^\times(v) \times H_w^\times(w) \\
    H(v,w) &= H_v^+(v) + H_w^+(w)    
  \end{aligned}
\end{equation}
after which (\ref{eq:elliptic_PDE}) takes the form
\begin{equation}
  \label{eq:substituted_PDE}
  \begin{aligned}
    0 &= H_w^\times \lbrack \frac{3}{v} (H_v^\times)^\prime +
    (H_v^\times)^{\prime\prime} \rbrack
    + e^{\Delta_{gf}-c_h} H_v^\times \lbrack \frac{1}{w}
    (H_w^\times)^\prime + (H_w^\times)^{\prime\prime} \rbrack \\
    0 &= \lbrack \frac{3}{v} (H_v^+)^\prime +
    (H_v^+)^{\prime\prime} \rbrack
    + e^{\Delta_{gf}-c_h} \lbrack \frac{1}{w}
    (H_w^+)^\prime + (H_w^+)^{\prime\prime} \rbrack
  \end{aligned}
\end{equation}
The crucial point is that, independently of $\Delta_{gf}(w)$ and
$c_h$, any solution of the ODEs
\begin{equation}
  \label{eq:separated_ODEs}
  \begin{aligned}
    H_v^{\prime\prime}(v) &= -\frac{3}{v} H_v^\prime(v) \\
    H_w^{\prime\prime}(w) &= -\frac{1}{w} H_w^\prime(w)
  \end{aligned}
\end{equation}
gives a solution of type IIB supergravity. Of course, finding an
analytic solution to (\ref{eq:separated_ODEs}) is quite
straightforward. As a matter of fact, these are the harmonic functions of
(\ref{eq:lesser_known_solutions})
\begin{equation}
  \begin{aligned}
    H_v &= \frac{c_{v1}}{v^2} + c_{v2} \\
    H_w &= c_{w2} \log w + c_{w3}
  \end{aligned}
\end{equation}
with $c_{v1}, c_{v2}, c_{w2}, c_{w3} \in \mathbb{R}$.
And so we have two new families of analytic solutions
\begin{equation}
  \label{eq:new_analytic_solution}
  \begin{aligned}
    H^\times(v,w) &= (c_{w2} \log w + c_{w3}) ( \frac{c_{v1}}{v^2} +
    c_{v2} ) \\
    H^+(v,w) &= c_{w2} \log w + \frac{c_{v1}}{v^2} + c_{v2} + c_{w3}
  \end{aligned}
\end{equation}
Of course these are just (\ref{eq:lesser_known_solutions}) and one
might ask what is new. The point is that
(\ref{eq:new_analytic_solution}) hold together with
(\ref{eq:analytic_metric}) for arbitrary $\Delta_{gf}(w)$, and hence
for arbitrary D$7$-brane distributions. The interpretation of these
solutions is similar to that given at the end of section
\ref{sec:known-solutions}. The D$3$-branes are smeared over some of
their transverse directions, but now also accomodate for any D$7$
distribution imposed by choice of $\Delta_{gf}(w)$.

It is interesting to note that (\ref{eq:lesser_known_solutions})
reappear as (\ref{eq:new_analytic_solution}) independently of whether
we add D$7$-sources or not. Of course, it would be much more
interesting to find the equivalent of (\ref{eq:D3-brane_solution}) in
the presence of $\Delta_{gf} \neq 0$. We shall do so in section
\ref{sec:numeric-solutions} numerically.

\subsection{Brane distributions}
\label{sec:brane-distributions}

At this point we will take a look at a few brane distributions. Note that
\begin{equation}
  \begin{aligned}
    \rho_{(2)} &= \lbrack (\partial_w^2 + \frac{1}{w} \partial_w)
    e^{-2\Delta_{gf}} \rbrack w \td w \wedge \td \phi \\
    &= (\square_w e^{-2\Delta_{gf}(w)}) w \td w \wedge \td\phi
  \end{aligned}
\end{equation}

From out ansatz it follows that D$7$-branes are always smeared along
$\phi$, so the simplest distribution is a $\delta$-function one in the
$w$ direction,
\begin{equation}
  \begin{aligned}
    \rho_{(2)} &= Q \delta(w-w_0) w \td w \wedge \td \phi
  \end{aligned}
\end{equation}
where $Q$ is some normalization constant. We can integrate the
resulting flux,
\begin{equation}
  \begin{aligned}
    \int_{S^1} F_{(1)} &= 2\pi Q w_0 \theta(w-w_0)
  \end{aligned}
\end{equation}
and it follows that
\begin{equation}
  \begin{aligned}
    Q = \frac{N_f}{2\pi w_0}
  \end{aligned}
\end{equation}
Then
\begin{equation}
  \label{eq:Delta_for_dirac_source}
  \begin{aligned}
    e^{-2\Delta_{gf}} &= \frac{N_f}{2\pi w_0} \lbrack c_1 \log w + w_0
    \theta (w-w0) \log \frac{w}{w0} + c_{\text{src}} \rbrack
  \end{aligned}
\end{equation}
Naturally this should be positive for all values of $w \in
\mathbb{R}^+$, hence it seems appropriate to set
$c_1 = 0$. Also, as both $e^{-\Delta_{gf}}$ and $e^{\Delta_{gf}}$
appear in the metric, $e^{-2\Delta_{gf}} \geq 0$ is a good assumption
that is guaranteed by fixing $c_{\text{src}} > 0$. Furthermore, our
numerical studies in section \ref{sec:numeric-solutions} will show
that varying $c_{\text{src}}$ does influence the form of the solutions rather
strongly. To avoid this, we will fix it to $c_{\text{src}} = Q^{-1}$
so that the constant term in $e^{-2\Delta_{gf}}$ does not vary with $N_f$.

A similarly interesting case is given by
\begin{equation}\label{eq:theta_brane_distribution}
  \begin{aligned}
    \rho_{(2)} &= Q \theta(w-w_0) w \td w \wedge \td\phi \\
    e^{-2\Delta_{gf}} &= Q \lbrack c_1 \log w + \frac{1}{4}
    \theta(w-w_0) (w^2 - w_0^2 - 2 w_0^2 \log \frac{w}{w_0}) +
    c_{\text{src}} \rbrack
  \end{aligned}
\end{equation}
For the same reasons as above we fix $c_1 = 0$ and
$c_{\text{src}} = Q^{-1}$. Concerning the normalization, we have
\begin{equation}
  \begin{aligned}
    F_{(1)} &= \frac{Q}{2} (w^2-w_0^2) \theta(w-w_0) \td \phi
  \end{aligned}
\end{equation}
leading to a radially dependent charge
\begin{equation}
  \begin{aligned}
    N_f(w) &= Q\pi (w^2 - w_0^2)
  \end{aligned}
\end{equation}
The fact that $N_f(w)$ behaves like a two-dimensional area is no
accident. After all, we assume a homogeneous brane distribution in the
$(w,\phi)$ plane for all $w \geq w_0$.

\subsection{Numeric solutions}
\label{sec:numeric-solutions}

Let us now take a look at numeric solutions of
(\ref{eq:elliptic_PDE}). We are dealing with a deformation of the
Laplace (Poisson) equation, that is, a homogeneous, elliptic,
separable PDE of second order, and use the Fortran package
Mudpack\footnote{Mudpack can be found at
  \href{http://www.cisl.ucar.edu/css/software/mudpack/}{http://www.cisl.ucar.edu/css/software/mudpack/}.
}
to do so. Our aim is to perform a qualitative study of deformations of
the original $AdS_5 \times S^5$ solution (\ref{eq:D3-brane_solution})
that includes additional D$7$-branes. We fix the parameter $r_3 = 1$
and solve the equation in a rectangular domain in the $(v,w)$
plane specified by 
\begin{equation}
  0.2 \leq v,w \leq 2.6
\end{equation}
on a $129 \times 129$ grid.
Some experimentation shows that one obtains a good agreement with the
analytic solutions in the absence of D$7$ branes when imposing the
Neumann boundary conditions at $w = 0.2$ and $w = 2.6$ and Dirichlet
ones at $v = 0.2$ and $v = 2.6$. I.e.~
\begin{equation}\label{eq:boundary_conditions}
  \begin{aligned}
    H &= \frac{1}{(v^2 + w^2)^2} & &\text{at } w = 0.2 \lor w = 2.6 \\
    \partial_v H &= -\frac{4 v}{(v^2 + w^2)^3} & &\text{at } v = 0.2
    \lor v = 2.6 \\
  \end{aligned}
\end{equation}
However, the physical significance of the boundary conditions is not
entirely clear and it might be appropriate to modify the boundary
conditions when changing the source density $\Delta_{gf}$.

\begin{figure}[tbhp]
  \centering
  \includegraphics{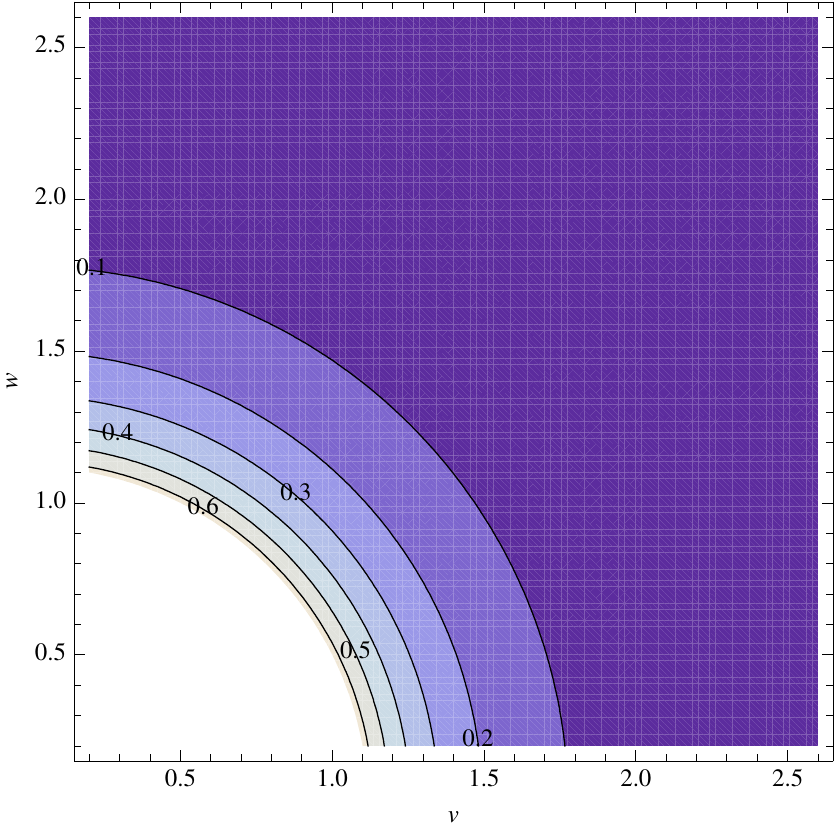}
  \caption{Plot of the analytic solution $H_{3} = (v^2 +
    w^2)^{-2}$. There is a singularity in $H_3$ at the origin
    charactereistic to the presence of D$3$ branes}
  \label{fig:ads5}
\end{figure}

\begin{figure}[hbtp]
  \centering
  \includegraphics{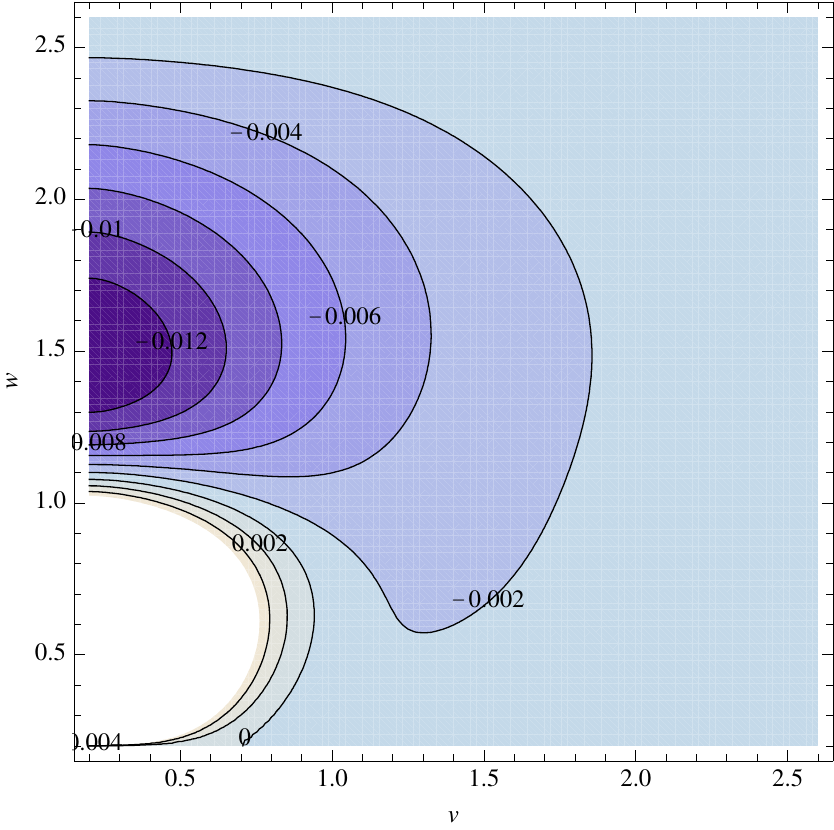}
  \caption{$\Delta H$ for $e^{-2\Delta_{gf}} = \theta( w
    - 1 ) \log ( w )+ 1$.}
  \label{fig:d7_times_1}
\end{figure}

\begin{figure}[hbtp]
  \centering
  \includegraphics{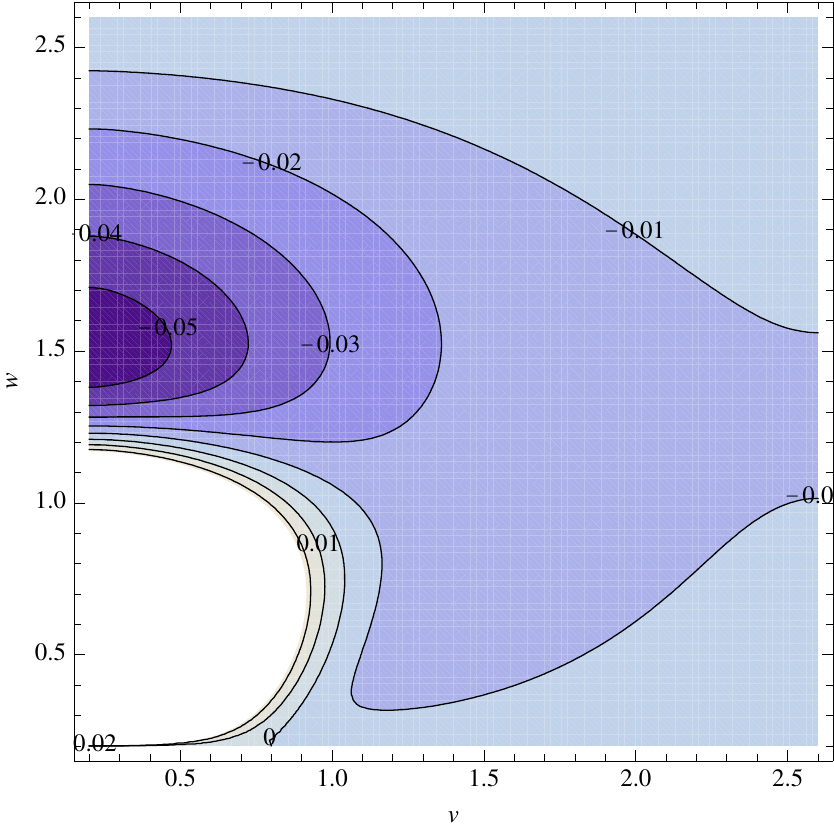}
  \caption{$\Delta H$ for $e^{-2\Delta_{gf}} = 10 \lbrack \theta( w
    - 1 ) \log ( w ) \rbrack + 1$.}
  \label{fig:d7_times_10}
\end{figure}

Figure \ref{fig:ads5} shows the analytic solution $H_{3} = \frac{1}{(v^2 +
  w^2)^2}$. Our numeric solution for $e^{-\Delta_{gf}} = 1$ (not
shown) agrees up to $\Delta H = \pm 0.0001$.
We then proceed to include D$7$ branes via changing
$e^{-2\Delta_{gf}}$. In all these cases we approximate Heaviside
$\theta$ functions by
\begin{equation}
  \label{eq:theta_function_approximation}
  \begin{aligned}
    \theta( w ) &= \frac{1}{2} + \frac{1}{2} \tanh (k w) \\
    k &= 2.5
  \end{aligned}
\end{equation}
Larger values of $k$ make for a sharper transition, in the case $k =
2.5$ we have $1-\theta(0.5) \sim 0.0758$. Figure \ref{fig:d7_times_1}
shows the case $e^{-2\Delta_{gf}} = \theta( w - 1 ) \log (
w ) + 1$ while figure \ref{fig:d7_times_10} uses
$e^{-2\Delta_{gf}} = 10 \lbrack \theta( w - 1 ) \log ( w )
\rbrack + 1$. So in each case, there is a stack of D$7$ branes
localized at $w = 1$, yet smeared along $\phi$. The changes in the
solutions are not drastic, but differ from $H_3$ by one or two orders
of magnitude, so instead of plotting $H$ for each case, we show the
difference to the pure D$3$-brane solution of figure \ref{fig:ads5},
$H - H_3$.

\begin{figure}[hbtp]
  \centering
  \includegraphics{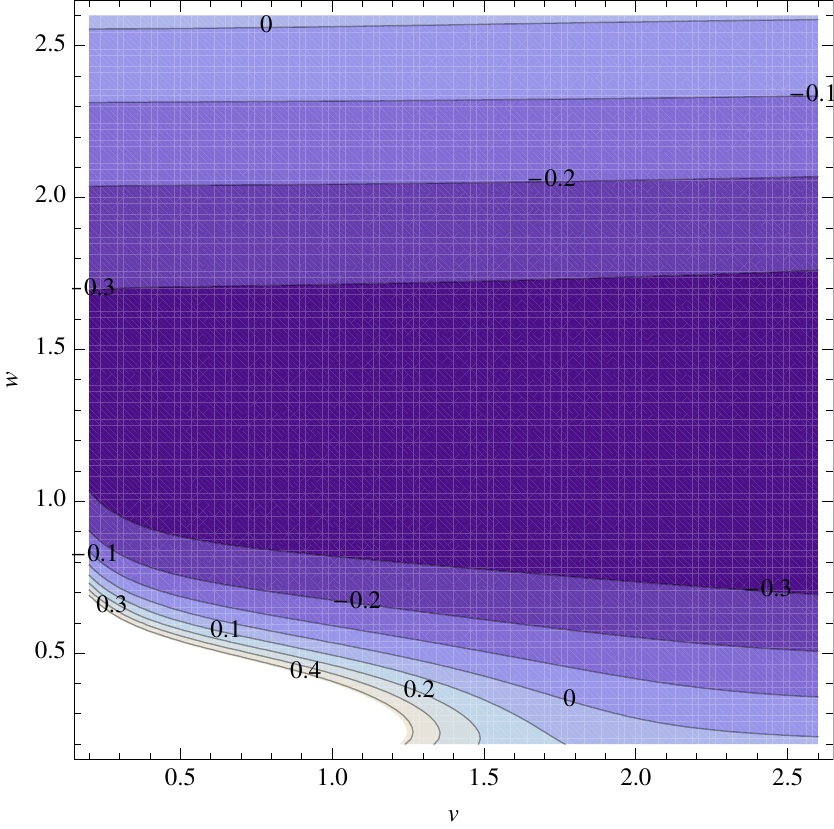}
  \caption{$H$ for $e^{-2\Delta_{gf}} = 500 \lbrack \theta( w
    - 1 ) \log ( w ) \rbrack + 1$.}
  \label{fig:d7_times_500}
\end{figure}

Things change considerably when we scale the source density by another
factor of $50$, i.e.~we set $e^{-2\Delta_{gf}} = 500 \lbrack \theta( w
    - 1 ) \log ( w ) + \rbrack 1$ (fig.~\ref{fig:d7_times_500}).
One can see quite clearly that the background is dominated by the D$7$
branes extending along the $v^i$ while the boundary conditions,
especially at $(0.2, 0.2)$ are still those of the D$3$ background.

\begin{figure}[hbtp]
  \centering
  \includegraphics{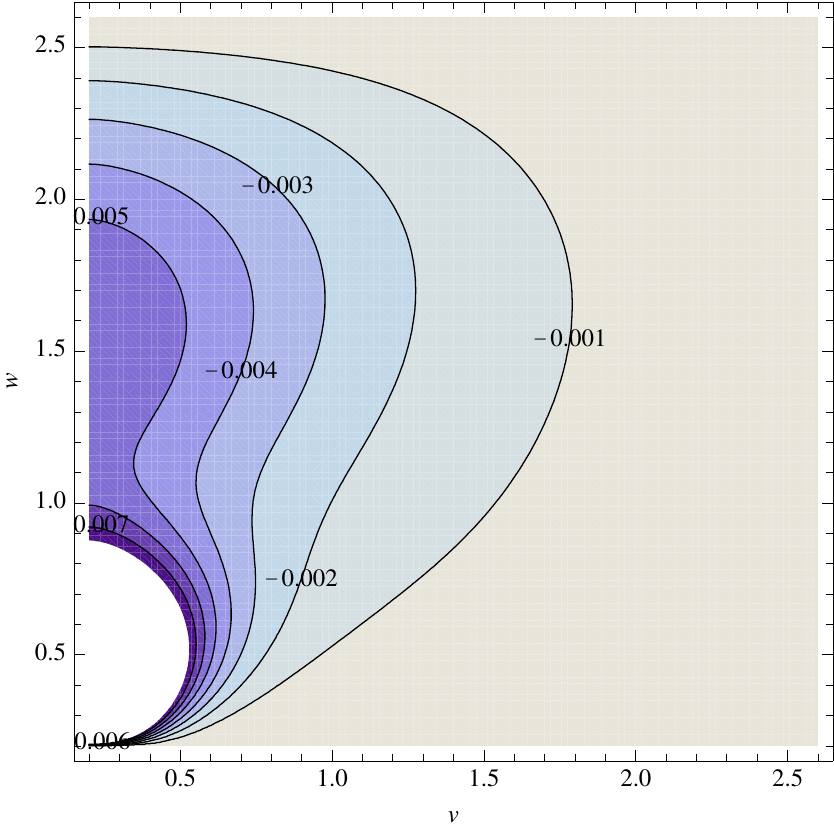}
  \caption{$\Delta H$ for $e^{-2\Delta_{gf}} = \frac{1}{4}
    \theta(w-1) \lbrack w^2 - 1- 2 w^2 \log(w)
    \rbrack + 1$.}
  \label{fig:d7_theta_times_1}
\end{figure}

\begin{figure}[bhtp]
  \centering
  \includegraphics{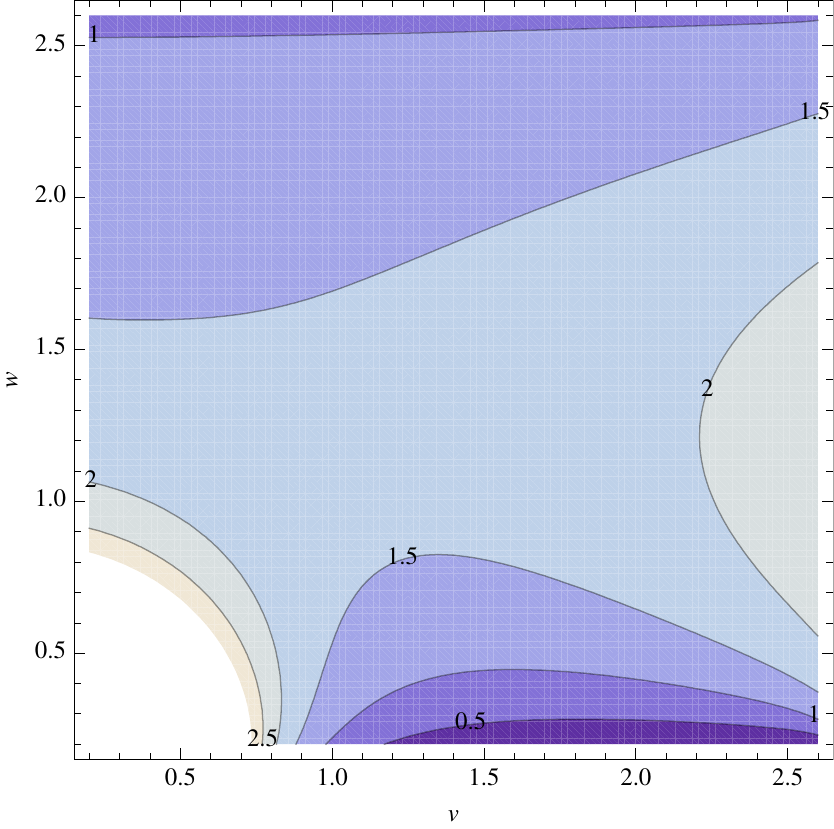}
  \caption{$H$ for $e^{-2\Delta_{gf}}$ as in
    fig~\ref{fig:d7_theta_times_1}, however, at $v = 2.6$ and $w =
    2.6$ we imposed the boundary conditions characteristic for D$3$
    branes smeared along $v^i$ -- a system T-dual to D$7$ branes.}
  \label{fig:d7_theta_bndy}
\end{figure}

Figure \ref{fig:d7_theta_times_1} shows a brane distribution along the
lines of (\ref{eq:theta_brane_distribution}). That is, the number of
flavors runs with $w^2$.
Of course, here the UV should be dominated by increasing number of
D$7$ branes and it might be appropriate to adjust the boundary
conditions at $v = 2.6$ and $w = 2.6$. Based on the T-dual of the
analytic D$7$ solution, we set them to
\begin{equation}
  \label{eq:modified_boundary_conditions}
  \begin{aligned}
    H &= \log w & &\text{at } w = 2.6 \\
    \partial_v H &= \frac{1}{w} & &\text{at } v = 2.6 \\
  \end{aligned}  
\end{equation}
while those at $v = 0.2$ and $w = 0.2$ remain as in
(\ref{eq:boundary_conditions}). The result is shown in
\ref{fig:d7_theta_bndy}. Note that having changed the boundary
conditions, the solution is quite different to \ref{fig:ads5} and we
plot $H$ instead of $\Delta H$.

\section{Conclusions}
\label{sec:conclusions}

In this paper, we have analyzed (\ref{eq:flavor_action}) from several
perspectives. From the perspective of the $p$-brane action in sections
\ref{sec:p-brane-sources} and \ref{sec:smeared-sources}, we realized
that color- and flavor-branes are actually on a very similar
footing. In principle one should include source terms for
both as was done in \cite{Bertolini:2001qa}, so this observation is
not new, as we mentioned before. However the presence of source terms
is only necessary if the associated sources are to be smeared over an
open subset of space-time, which is why the sourc-term is essential
for the flavor-branes that are usually assumed to be smeared.

The impression that color and flavor-branes can -- from the supergravity
perspective -- be taken to be on an equal footing was again confirmed
by our observations in section \ref{sec:an-aside-t-duality}, where we
were able to exchange color and flavor-branes by performing four
T-dualities in the directions transverse to the D$3$s. Curiously, we
had included explicit source terms for the (smeared) flavor D$7$-branes while
not doing so for their localized\footnote{
As a matter of fact, the D$3$s were smeared over some of their
transverse directions in the analytic solutions presented
in~\ref{sec:analytic-solutions}, yet they were not smeared over an
open subset of space-time.} 
cousins. One should also keep in mind that \cite{Kirsch:2005uy}
obtained highly similar results working with the supergravity action
alone -- while including suitable $\delta$-function sources, of
course.

We also raised the issue whether it is generally possible to find a
source-term for a given solution -- especially in cases where
supersymmetry is broken. As discussed in \cite{Ortin:2004ms}, the
problem lies in the fact that for sources in theories of gravity, the
energy of the source is not localized at the source but also stored in
the self-energy of the surrounding gravitational field. Only in the
presence of supersymmetry, where gravitational effects are canceled by
those of a different field -- the Maxwell-type $p$-form fields in this
case -- can one find a suitable source term. Again we point out
\cite{Bigazzi:2009bk} however, where the authors have constructed a
finite-temperature background including additional flavor terms.

Naturally our comments and observations made here are only valid for
the examples studied, and it would be interesting to study the issue
of source terms for color-branes for more complex supergravity
backgrounds dual to confining gauge theories, such as
\cite{Maldacena:2000yy}, \cite{Klebanov:2000hb} and
\cite{Polchinski:2000uf}. From the point of gauge/string duality, the
crucial point is there whether there are open string states in the
spectrum, that should only appear in non-confining theories. In other
words, one expects that for confining backgrounds it should not be
possible to find source terms for the color branes. Verifying this
explicitly would be an avenue item for future research.

Finally, we found a series of new D$3$-D$7$ backgrounds with smeared
D$7$-branes. Here, the analytic solutions captured in
(\ref{eq:new_analytic_solution}) have the interesting property that
for any distribution of D$7$-branes encoded in $\Delta_{gf}$, the
D$3$s distribute themselves such there is a solution. It would be
intersting to interpret the solutions of sections
\ref{sec:analytic-solutions} and \ref{sec:numeric-solutions} in the
context of gauge/string duality, another problem that we leave for
future work.

\section*{Acknowledgements}
\label{sec:acknowledgements}

I would like to thank Carlos N\'u\~nez, Prem Kumar,
Tim Hollowood, Stefan Hollands, \'Angel Paredes and Daniel Arean for
useful discussions as well as comments on the manuscript. My work is
supported by the German National Academic Foundation (Studienstiftung
des deutschen Volkes) as well as an STFC studentship.

\bibliographystyle{plain}

\begin{thebibliography}{99}
\bibitem{Bigazzi:2005md}
 F.~Bigazzi, R.~Casero, A.~L.~Cotrone, E.~Kiritsis and A.~Paredes,
 JHEP {\bf 0510}, 012 (2005)
 [arXiv:hep-th/0505140].

\bibitem{Casero:2006pt}
 R.~Casero, C.~Nunez and A.~Paredes,
 Phys.\ Rev.\  D {\bf 73}, 086005 (2006)
 [arXiv:hep-th/0602027].

\bibitem{Nunez:2010sf}
 C.~Nunez, A.~Paredes and A.~V.~Ramallo,
 arXiv:1002.1088 [hep-th].

\bibitem{Maldacena:2000yy}
 J.~M.~Maldacena and C.~Nunez,
 Phys.\ Rev.\ Lett.\  {\bf 86}, 588 (2001)
 [arXiv:hep-th/0008001].

\bibitem{Klebanov:1999tb}
 I.~R.~Klebanov and E.~Witten,
 Nucl.\ Phys.\  B {\bf 556}, 89 (1999)
 [arXiv:hep-th/9905104].

\bibitem{Bertolini:2000dk}
 M.~Bertolini, P.~Di Vecchia, M.~Frau, A.~Lerda, R.~Marotta and I.~Pesando,
 JHEP {\bf 0102}, 014 (2001)
 [arXiv:hep-th/0011077].

\bibitem{Bertolini:2001qa}
 M.~Bertolini, P.~Di Vecchia, M.~Frau, A.~Lerda and R.~Marotta,
 Nucl.\ Phys.\  B {\bf 621}, 157 (2002)
 [arXiv:hep-th/0107057].

\bibitem{Aharony:1998xz}
 O.~Aharony, A.~Fayyazuddin and J.~M.~Maldacena,
 JHEP {\bf 9807}, 013 (1998)
 [arXiv:hep-th/9806159].

\bibitem{Grana:2001xn}
 M.~Grana and J.~Polchinski,
 Phys.\ Rev.\  D {\bf 65}, 126005 (2002)
 [arXiv:hep-th/0106014].

\bibitem{Kirsch:2005uy}
 I.~Kirsch and D.~Vaman,
 Phys.\ Rev.\  D {\bf 72}, 026007 (2005)
 [arXiv:hep-th/0505164].

\bibitem{Ortin:2004ms}
 T.~Ortin,
 ``Gravity And Strings,''
{\it  Cambridge Unversity, Cambridge University Press, 2004}

\bibitem{Klebanov:2000hb}
 I.~R.~Klebanov and M.~J.~Strassler,
 JHEP {\bf 0008}, 052 (2000)
 [arXiv:hep-th/0007191].

\bibitem{Bigazzi:2009bk}
 F.~Bigazzi, A.~L.~Cotrone, J.~Mas, A.~Paredes, A.~V.~Ramallo and J.~Tarrio,
 JHEP {\bf 0911}, 117 (2009)
 [arXiv:0909.2865 [hep-th]].

\bibitem{Evans:1998}
  L.~C.~Evans,
  ``Partial Differential Equations,''
  American Mathematical Society,
  1998

\bibitem{Casero:2007jj}
 R.~Casero, C.~Nunez and A.~Paredes,
 Phys.\ Rev.\  D {\bf 77}, 046003 (2008)
 [arXiv:0709.3421 [hep-th]].

\bibitem{HoyosBadajoz:2008fw}
 C.~Hoyos-Badajoz, C.~Nunez and I.~Papadimitriou,
 Phys.\ Rev.\  D {\bf 78}, 086005 (2008)
 [arXiv:0807.3039 [hep-th]].

\bibitem{Benini:2006hh}
 F.~Benini, F.~Canoura, S.~Cremonesi, C.~Nunez and A.~V.~Ramallo,
 JHEP {\bf 0702}, 090 (2007)
 [arXiv:hep-th/0612118].

\bibitem{Bigazzi:2008qq}
 F.~Bigazzi, A.~L.~Cotrone, A.~Paredes and A.~V.~Ramallo,
 JHEP {\bf 0903}, 153 (2009)
 [arXiv:0812.3399 [hep-th]].

\bibitem{Gaillard:2008wt}
 J.~Gaillard and J.~Schmude,
 JHEP {\bf 0901}, 079 (2009)
 [arXiv:0811.3646 [hep-th]].

\bibitem{Gaillard:2009kz}
 J.~Gaillard and J.~Schmude,
 JHEP {\bf 1002}, 032 (2010)
 [arXiv:0908.0305 [hep-th]].

\bibitem{Benini:2007gx}
 F.~Benini, F.~Canoura, S.~Cremonesi, C.~Nunez and A.~V.~Ramallo,
 JHEP {\bf 0709}, 109 (2007)
 [arXiv:0706.1238 [hep-th]].

\bibitem{Bigazzi:2008zt}
 F.~Bigazzi, A.~L.~Cotrone and A.~Paredes,
 JHEP {\bf 0809}, 048 (2008)
 [arXiv:0807.0298 [hep-th]].

\bibitem{Arean:2008az}
 D.~Arean, P.~Merlatti, C.~Nunez and A.~V.~Ramallo,
 JHEP {\bf 0812}, 054 (2008)
 [arXiv:0810.1053 [hep-th]].

\bibitem{Koerber:2007hd}
 P.~Koerber and D.~Tsimpis,
 JHEP {\bf 0708}, 082 (2007)
 [arXiv:0706.1244 [hep-th]].

\bibitem{Polchinski:2000uf}
  J.~Polchinski and M.~J.~Strassler,
  arXiv:hep-th/0003136.
\end{thebibliography}

\end{document}